\documentclass{article}
\usepackage{amsmath, amssymb, graphicx}
\usepackage{subfigure}

\newtheorem{theorem}{Theorem}

\newtheorem{corollary}[theorem]{Corollary}

\title{A Mean-Field Theory of Cellular Automata Model for Distributed
Packet Networks}

\author{Maoke CHEN \and Tao HE \and Xing LI}

\begin{document}
\pagestyle{plain} \maketitle
\begin{abstract}
    A Mean-Field theory is presented and applied to a Cellular
    Automata model of distributed packet-switched networks. It is
    proved that, under a certain set of assumptions, the critical
    input traffic is inversely proportional to the free packet delay
    of the model. The applicability of Mean-Field theory in queue
    length estimation is also investigated. Results of theoretical
    derivations are compared with simulation samples to demonstrate
    the availability of the Mean-Field approach.
\end{abstract}

\section{Introduction}

Modelling computer networks is important for understanding the
network behaviors, especially those related to the critical
phenomena. Models must assign an assumption to network topology.
Some of them make the assumption based on graph theory
\cite{trtyakov98phase}, while some others based on regular
lattices
\cite{deane96selfsimilarity,fuks99performance,liu02simple,ohira98phase}.
What kind of models is chosen depends upon the application
background.

This paper focuses on architectures of distributed packet-switched
networks. Some overlays, mobile \emph{ad-hoc} networks and
metropolitan area networks fall into the category of
``distributed'', without any central or hierarchical control and
without differentiation among transit and end-user sites. On
account of these characteristics, the model presented in this
paper is extended from Fuk\'s' work \cite{fuks99performance}.
Actually, the same topology was studied even in the pre-Internet
history, within the reports of RAND on distributed networks
authored by Paul Baran and others
\cite{baran64history,baran64introduction}. However, only
survivability of the lattices was studied there. We step forward
this effort into dynamic behavior of packet-switched Cellular
Automata over the lattices.

On the other hand, previous works on Cellular Automata for data
networks have discovered in simulations that critical traffic
behavior is related to free packet delay in the networks
\cite{fuks99performance}. By the help of Mean-Field Theory
technique, we step forward this discovery to an approximated
analytical theorem for the extended model, where the critical
traffic is inversely proportional to the free delay of the
network, provided the assumptions of Mean-Field Theory are
available. The most basic idea in the approximation relies on
simplifying the system with an identical open Jackson network.
However, the application of the Mean-Field Theory could not be
exaggerated. In estimating queue length of the model, the
Mean-Field approach is not accurate.

The rest parts of the paper are organized as such: section 2
describes the model and its parameters. Section 3 applies
Mean-Field Theory to the model, with a certain set of
approximations. Some simulation results are shown, in order to
demonstrate that the Mean-Field Theory presented here are
available for cases with any dimensionality and any neighborhood.
Section 4 briefly discusses queue length estimated by the
Mean-Field Theory, with comparison to simulation data. Finally, we
summarize the works with emphasizing its significance to analysis
and design of distributed network architectures.

\section{Model Definitions}

Cellular Automaton is a mathematical model for physical systems
containing large amount of simple, identical and locally
interacting units \cite{wolfram84uni}. Any Cellular Automata could
be defined with a 4-tuple of lattice space, neighborhood, state
set, and rule of state-transition \cite{weimar95cellular}. The
Cellular Automata models for distributed packet-switched networks
(briefly ``the model'' or ``our model'', later through the paper)
are also defined as such.

\subsection{Lattice}

The model are defined on $d$-dimensional Euclidean lattice space.
Originally the lattice is boundless and extended to infinity. For
digital simulation, however, the lattice is often truncated in a
certain $d$-dimensional hypercube, say $L$ as its width. Because a
distributed network has no geometrical center, the truncated
lattice should be thought as periodical, i.e. the coordinate
values with same remainder modulo $L$ are identical. Therefore the
lattice of the model is denoted with
\begin{equation}\label{eqn:latticeDefinition}
    \mathcal{L}^d \triangleq \mathbb{Z}^d \cap [0, L)^d
\end{equation}

And the bases of the lattice space are denoted with $\mathbf{e}_i,
i = 1, 2, \cdot, d$.

\subsection{Neighborhood and Metric}

A neighborhood is a mapping from the lattice to its power set:
\begin{eqnarray}\label{eqn:neighborhoodDefinition}
    A &:& \mathcal{L}^d \mapsto P(\mathcal{L}^d)
\end{eqnarray}
For the purpose of routing packets among the sites in the model,
metrics are defined with the neighborhood as well. For example,
von Neumann neighborhood and the periodic Taxicab metric
\footnote{Or Manhattan metric.} are defined by:
\begin{eqnarray}\label{eqn:taxicabDefinition}
    A(\mathbf{x}) &=& \bigcup_{i=1}^d \{\mathbf{x} + \mathbf{e}_i, x -
        \mathbf{e}_i\}, \quad \forall \mathbf{x} \in \mathcal{L}_d
        \label{eqn:taxicabDefinition:neighborhood}\\
    D(\mathbf{x}, \mathbf{y}) &=& \sum_{i=1}^d
        {\frac{L}{2}-\left|\left|x_i - y_i\right|-\frac{L}{2}\right|}
        \label{eqn:taxicabDefinition:metric}
\end{eqnarray}

While Moore neighborhood and the periodic Moore metric is defined
as follows:
\begin{eqnarray}\label{eqn:mooreDefinition}
    A(\mathbf{x}) &=& \left(\bigcup_{y_1=-1,0,1}{\cdots\bigcup_{y_d=-1,0,1}
        {\left\{\mathbf{x} + \sum_{i=1}^d{y_i \mathbf{e}_i}\right\}}}\right)
        \backslash\{\mathbf{x}\} \label{eqn:mooreDefinition:neighborhood}\\
    D(\mathbf{x}, \mathbf{y}) &=& \max_{i}
        \left\{\frac{L}{2}-\left|\left|x_i - y_i\right|-\frac{L}{2}\right|\right\}
        \label{eqn:mooreDefinition:metric}
\end{eqnarray}

Definitions in both (\ref{eqn:taxicabDefinition:metric}) and
(\ref{eqn:mooreDefinition:metric}) conforms to the lattice
periodicity. The two metrics are identical in 1-dimensional cases.

\subsection{State and Transition}

The state of a site is a first-in-first-out queue of packets, each
of which contains at least the information about its destination.
Each site $\mathbf{x}$'s queue length at time $k$, denoted as
$q(\mathbf{x}, k)$, is modified by both packet input and packet
forwarding processes.

At each discrete moment of time $k$, packet enters the network
from outside any site independently with an identical probability
$\lambda$. The probability of more than one input packets is
$o(\lambda)$. Thus, one can think there is a discrete-time Poisson
source at each site. Destination of a newly entering packet is
randomly selected among all possible sites with the same
probability. A new packet is queued at the site where it enters
the network.

At each moment, a site serves the first packet in its queue,
forwarding it to one of its neighbors properly selected with the
routing criteria. The service time is a constant of unity. Two
criteria are applied to route selection. First, the next-hop
should be selected from the neighbors, nearest to the destination
in the term of given metric. The nearest neighbor set of a site
$\mathbf{x}$ to destination $\mathbf{z}$ is
\begin{equation}\label{eqn:minimumDistanceCriterion}
    B(\mathbf{z}; \mathbf{x}) = \{\mathbf{y} \in A(\mathbf{x}):
    D(\mathbf{z}, \mathbf{y}) \rightarrow \min\}
\end{equation}

Second, the next-hop should be selected from the neighbors with
minimum queue length within the neighbors nearest to destination.
\begin{equation}\label{eqn:minimumQueueing}
    C(\mathbf{z}; \mathbf{x}, k) = \{\mathbf{y} \in B(\mathbf{z}; \mathbf{x}):
    q(\mathbf{y}, k)\rightarrow \min\}
\end{equation}

Finally, if the minimum-queue nearest-to-destination neighbor set
$C(\mathbf{z}; \mathbf{x}, k)$ contains more than one coordinates,
then anyone is selected as the next-hop, randomly with the same
probability. If this one is current destination $\mathbf{z}$, then
the packet is not queued anymore: it leaves the system.

Baran's distributed network model contains various neighborhoods
\cite{baran64history}. To extend them into higher dimensionality,
von Neumann and Moore neighborhoods are definitely well defined
while other neighborhoods might not. Fuk\'s' two-dimensional von
Neumann Cellular Automata model allows non-periodical property
\cite{fuks99performance}, but it's less close to the case of
distributed networks where all the sites are identical in
topology.

\section{Mean-Field Theory}

Any queueing system has a critical traffic as the upper bound of
the input traffic, so that the system converges into a stable
state instead of going far from stability. This critical traffic
is just the service rate $\mu$ in a single queueing-service
system, while it is different for networks.

Ohira has shown that the critical traffic is related to the free
delay of the network \cite{ohira98phase} and Fuk\'s has uncovered
that, in two-dimensional von Neumann Cellular Automata model, the
sufficient and necessary condition for stability of the model is
$\lambda < 1 / \bar\tau_0$, or equivalently,
\begin{equation}\label{eqn:criticalTrafficLaw}
    \lambda_c = \frac{1}{\bar\tau_0}
\end{equation}
where $\bar\tau_0$ is the average delay of a free packet without
being queued anywhere. These previous results are observed in
simulations. Now we prove that the law of
(\ref{eqn:criticalTrafficLaw}) is an analytical result under
certain assumptions and is available for whatever dimensionality
and neighborhoods rather than only for two-dimensional von Neumann
cases.

\subsection{Approximation to Open Jackson Network}

The model defined in the previous section has the property of open
queueing networks, i.e. all packets enter the system from outside
and finally leave at the destination. This inspires utilizing
ever-known conclusions in open Jackson network.

Open Jackson network is of Markovian queueing network, i.e. packet
arrival is a Poisson process and the service time of each site
conforms to exponential distribution. The Jackson theorem presents
the condition of stability as well as the queue length
distribution in stable state \cite{kleinrock76queue,sheng98queue}.
In the Jackson theorem, a parameter $\sigma(\mathbf{x})$, called
as ``site traffic'', is defined as the traffic observed at site
$\mathbf{x}$ in the system and we have the traffic equilibrium
equations
\begin{equation}\label{eqn:jacksonTheorem:sigma}
    \sigma(\mathbf{x}) = \lambda(\mathbf{x}) +
    \sum_{\mathbf{y}}{\sigma(\mathbf{y}) r_{\mathbf{yx}}}, \quad \forall
    \mathbf{x}
\end{equation}
where $r_{\mathbf{yx}}$ is the forwarding probability from site
$\mathbf{y}$ to $\mathbf{x}$ and
\begin{equation}
    \sum_\mathbf{x}{r_{\mathbf{yx}}} = 1 - r_{\mathbf{y},\infty},
    \quad \forall \mathbf{y}
\end{equation}
where $r_{\mathbf{y},\infty}$ represents the probability of
leaving.

Our model, however, contains sites over a lattice space, each of
which is a discrete-time M/D/1 queueing system. Unfortunately, to
the best knowledge of the authors, there are hardly few approaches
of queueing networks consisting of M/D/1 systems. Then the
constant service time is replaced by a exponential distribution
with the mean value of $1/\mu = 1$ . Furthermore, we ignore the
details in route selection and approximate it with a simple,
time-invariant, and destination-free probability,
$r_{\mathbf{yx}}$,

Thus, the model is approximated with a continuous-time open
Jackson network over the lattice space, where input traffic is an
identical Poisson process at each site with parameter
$\lambda(\mathbf{x}) = \lambda$, and each site serves the traffic
with parameter $\mu = 1$.

\subsection{Mean-Field Theory for Critical Traffic}

\begin{figure}
  \centering
  \includegraphics[width=.8 \textwidth]{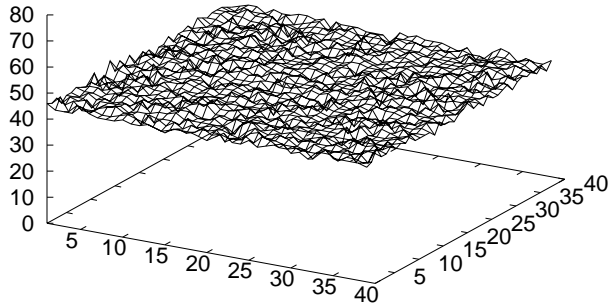}\\
  \caption{Queue length field of a sample with $d = 2, L = 40,
  \lambda = 0.08, k = 1000$, von Neumann neighborhood, and periodic
  taxicab metric. Queue lengths are so close to each
  other among all the sites that assuming they have an identical
  $\sigma$ is reasonable.}
  \label{fig:qFieldCQN}
\end{figure}

Observing samples of queue growth processes in the model (Figure
\ref{fig:qFieldCQN}), one can easily summarized that site traffic
$\sigma(\mathbf{x})$ seems to be a constant without difference
referring to the coordinates. This implies that a Mean-Field
Theory can be developed in order to get the critical traffic law
of the open Jackson network derived from the model.

Mean-Field Theory is an approximate technique widely used in
statistical physics, which treats the order-parameter as spatially
constant \cite{chaikin97principles}. For our model, the Mean-Field
approximation aims at an identical parameter $\sigma$ such that
\begin{equation}\label{eqn:MFT:sigma}
    \sigma(\mathbf{x}) = \sigma, \quad \forall \mathbf{x} \in
    \mathcal{L}^d
\end{equation}

Three heuristic conditions, called as Mean-Field Theory
assumptions, support this approximation.
\begin{description}
    \item[1. Isotropy] Lattice space, either infinite or periodic,
    is isotropic in geometry. Therefore, it is reasonable to assume
    that, at any moment, any site forwards the first packet with a
    same probability to its neighbors. That is
    \begin{eqnarray}\label{eqn:isotropy}
        r_{\mathbf{yx}} &=& r_{\mathbf{y}} \nonumber \\
        &=& \frac{1}{|A(\mathbf{y})|}
        \left(1 - r_{\mathbf{y},\infty}\right),
        \quad \forall \mathbf{x} \in
        A(\mathbf{y}), \quad \forall \mathbf{y} \in \mathcal{L}^d
    \end{eqnarray}
    \item[2. Homogeneity] Further, it is assumed that the forwarding
    probability are not different among all the sites. In addition
    to the formula (\ref{eqn:isotropy}) and the fact
    $|A(\mathbf{y})| \equiv A, \quad \forall \mathbf{y} \in \mathcal{L}^d$, we have
    \begin{eqnarray}\label{eqn:homogeneity}
        r_{\mathbf{y}} &=& r = \frac{1}{A}
        \left(1 - r_{\infty}\right) \nonumber \\[.5em]
        r_{\mathbf{y},\infty} &=& r_{\infty},
        \quad \forall \mathbf{y} \in \mathcal{L}^d
    \end{eqnarray}
    \item[3. Spatial and temporal ergodicity] It is assumed that
    the queue length process $q(\mathbf{x},k)$ is spatially ergodic, i.e.
    \begin{equation}\label{eqn:ergodicitySpatial}
        E\{q(\mathbf{x},k)\} = \lim_{L \rightarrow \infty}
        \frac{1}{|\mathcal{L}^d|}{\sum_{\mathbf{x} \in \mathcal{L}^d}
        {q(\mathbf{x},k)}}, a.s.
    \end{equation}
    And furthermore, it is also temporally ergodic as long as the
    stable state $q(\mathbf{x}) \triangleq \lim_{k \rightarrow \infty}
    {q(\mathbf{x}, k)}$ is achieved. Later on, we denote $\bar{q}$
    for either the ensemble, or temporal, or spatial average (in the stable
    state if it exists) of queue length processes on sites over the lattice.
\end{description}

From (\ref{eqn:isotropy}) and (\ref{eqn:homogeneity}), note that
$\lambda(\mathbf{x})$ are identical to $\lambda$, the traffic
equations (\ref{eqn:jacksonTheorem:sigma}) is simplified to
\begin{equation}\label{eqn:sigmaSimplified}
    \sigma(\mathbf{x}) = \lambda + \left[\frac{1}{A}(1 - r_{\infty})\right]
    \sum_{\mathbf{y}}\sigma(\mathbf{y}), \quad \forall \mathbf{x} \in
    \mathcal{L}^d
\end{equation}

For linear equations (\ref{eqn:sigmaSimplified}) are symmetric to
permutations of $\{\sigma(\mathbf{x})\}$, and the number of
equations are equal to the number of unknown variables, it is
definite that they have a unique solution set which is satisfying
(\ref{eqn:MFT:sigma}).

Then equation (\ref{eqn:sigmaSimplified}) are reduced to a single
equation referring to $\sigma$:
\begin{equation}\label{eqn:sigmaIdentical}
    \sigma = \lambda + (1 - r_{\infty}) \sigma
\end{equation}

and finally it is solved that
\begin{equation}
    \sigma = \frac{\lambda}{r_{\infty}}
\end{equation}

The value of $r_{\infty}$ has not yet determined. We'd like to
present the Mean-Field version of Jackson Theorem first and then
derive $r_{\infty}$ by applying the well-known Little's Law to the
mean value of packet lifetime.

With help of the assumptions above, and the Jackson Theorem, we
have
\begin{theorem}\label{theorem:MFTJackson}
    Under the assumptions of Mean-Field Thoery, the
    queueing network of the model converges to stable state if and
    only if
    \begin{equation}\label{eqn:MFTJackson:iff}
        \rho \triangleq \frac{\sigma}{\mu} = \frac{\lambda}{r_{\infty}}
        < 1
    \end{equation}
    And in the stable state, queue length on each site, $q(\mathbf{x})$,
    conforms to an identical geometric distribution:
    \begin{equation}\label{eqn:MFTJackson:prob}
        P(q(\mathbf{x}) = n) = (1 - \rho)\rho^n, \quad \forall n \ge
        0, \quad \forall \mathbf{x} \in \mathcal{L}^d
    \end{equation}
    And the mathematical expectation of stable queue length is:
    \begin{equation}\label{eqn:MFTJackson:qLength}
        \bar{q} = \frac{\rho}{1 - \rho}, \quad \forall \mathbf{x}
        \in \mathcal{L}^d
    \end{equation}
\end{theorem}

Now from the Little's Law of arbitrary queueing system, it holds
in the stable state that
\begin{equation}\label{eqn:littleLaw}
    \bar{q} = \lambda \bar\tau
\end{equation}

where $\bar\tau$ is the mathematical expectation of packet
lifetime in stable state. When all the sites have identical queue
length, lifetime of a packet is independent upon the path that it
pass through, and is equal to the free delay $\bar\tau_0$ plus the
total time of being queued. Note that the packet must be queued
$\bar\tau_0$ times, and apply the third assumption to regard queue
length of each site on a packet path is constant and equal to the
average $\bar{q}$, then we have
\begin{equation}\label{eqn:lifetime}
    \bar\tau = \bar\tau_0 + \bar\tau_0 \bar{q}
\end{equation}

Apply (\ref{eqn:lifetime}) to (\ref{eqn:littleLaw}), we obtain an
equation
\begin{eqnarray}\label{eqn:littleLawEquation}
    \bar{q} &=& \lambda (\bar\tau_0 + \bar\tau_0 \bar{q})
    \nonumber \\
    \text{and~~~} \bar{q} &=& \frac{\lambda \bar\tau_0}{1 -
    \lambda \bar\tau_0}
\end{eqnarray}

Recall the formulae (\ref{eqn:MFTJackson:iff}),
(\ref{eqn:MFTJackson:qLength}) in the Theorem
\ref{theorem:MFTJackson} and compare them to the equation
(\ref{eqn:littleLawEquation}), we have the following corollary,
which represents the law of critical traffic as a function of the
free delay in the model\footnote{The third assumption plays an
important role here. Note that in formula
(\ref{eqn:MFTJackson:qLength}), $\bar{q}$ is, in fact, the
ensemble average; while in Little's Law
(\ref{eqn:littleLawEquation}), $\bar{q}$ is actually a time
average. Without the ergodicity assumption, the two formula could
not be combined.}.

\begin{corollary}\label{theorem:MFTJackson:corollary}
    In the Mean-Field Theory of the model, the packet's leaving
    probability and the site traffic are respectively
    \begin{eqnarray}
        r_{\infty} &=& \frac{1}{\bar\tau_0}
        \label{eqn:MFTJackson:corollary:exitProb}\\[.5em]
        \text{and~~~}\sigma &=& \lambda \bar\tau_0
        \label{eqn:MFTJackson:corollary:sigma}
    \end{eqnarray}
    And the stability condition is equivalent to
    \begin{equation}\label{eqn:MFTJackson:corollary:criticalTraffic}
        \lambda < \frac{1}{\bar\tau_0}
    \end{equation}
    or equally to say the critical input traffic is
    $\lambda_c = \frac{1}{\bar\tau_0}$.
\end{corollary}

The corollary, esp. the formula
(\ref{eqn:MFTJackson:corollary:sigma}), shows that the free delay
of a network does significantly impact on the critical traffic.
Actually, delay \emph{amplifyings} input traffic to site traffic
linearly; or equivalently to say, free delay in a network decrease
the service capability of the sites.

It is emphasized that, in the Mean-Field Theory demonstrated
above, there is not any requirement to lattice dimensionality, nor
its neighborhood type. The analytical result of Theorem
\ref{theorem:MFTJackson} and its Corollary is universally
available, provided the Mean-Field Theory assumptions are
conforming to physical properties of the model.

\subsection{Simulation Samples}

\begin{figure}[htb]
  \centering
  \subfigure[$d = 1$, $L = 100$, $\bar\tau_0 = 25$,
  $\lambda_c = 0.04$]{
    \label{fig:profileQ:1dL100}
    \includegraphics[width=.45\textwidth]{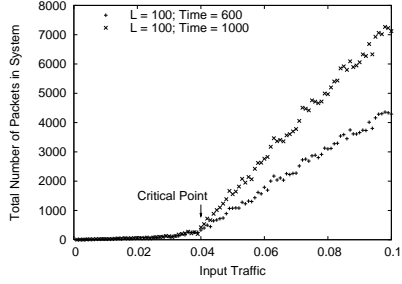}
  }\hfil
  \subfigure[$d = 1$, $L = 200$, $\bar\tau_0 = 50$,
  $\lambda_c = 0.02$]{
    \label{fig:profileQ:1dL200}
    \includegraphics[width=.45\textwidth]{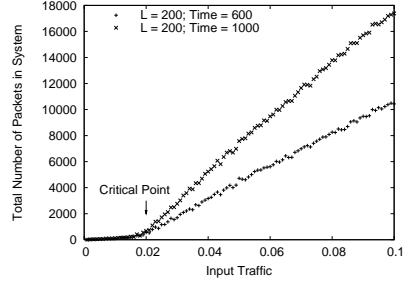}
  }\\
  \subfigure[von Neumann lattice,
  $d = 2$, $L = 40$, $\bar\tau_0 = 20$, $\lambda_c = 0.05$]{
    \label{fig:profileQ:2dNeumann}
    \includegraphics[width=.45\textwidth]{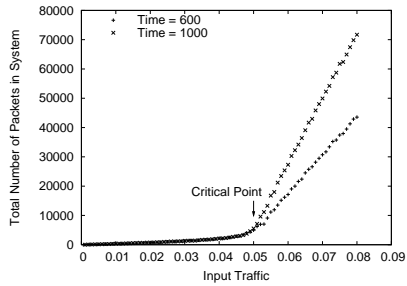}
  }\hfil
  \subfigure[Moore lattice,
  $d = 2$, $L = 30$, $\bar\tau_0 = \frac{1801}{180} \simeq 10.0$,
  $\lambda_c = 0.10$]{
    \label{fig:profileQ:2dMoore}
    \includegraphics[width=.45\textwidth]{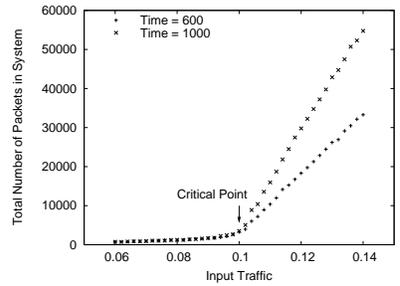}
  }
  \caption{Simulation samples to show the critical behavior. With
  whatever dimensionality and whatever neighborhood, the model's
  critical traffic $\lambda_c$ is almost equal to $1/\bar\tau_0$.}
  \label{fig:profileQ}
\end{figure}

As the upper bound of input traffic for existence of stable state,
the critical behavior in an instance of the model may be
demonstrated by its queue length processes or the total number of
packets in the entire system, say $Q(k)\triangleq
\sum_{\mathbf{x}}{q(\mathbf{x},k)}$. When the running time $k$ is
large enough, the model is thought of being stable if the total
number $Q(k)$ is almost unchanged as $k$ goes larger. If the input
traffic exceeds the critical point, then $Q(k)$ will increase
steadily to infinity. These phenomena are shown in Figure
\ref{fig:profileQ}, where several sample cases with variety of
parameters or topology characteristics are provided. Figure
\ref{fig:profileQ:2dNeumann} is similar to the case that
\cite{fuks99performance} has provided. However, we have extended
the results on critical traffic to far more general environments.

\section{Queue Length Estimation}

As any other Mean-Field approaches for statistical-physical
systems, the Mean-Field Theory presented here can \emph{not} be
treated as an accurate quantitative result, though it almost
accurately predicts the critical behavior of the model.

\subsection{Queue Length of Stable State}

Theorem \ref{theorem:MFTJackson} has given the mean value for the
queue length on any site in the stable state, if it exists.
However, the result overestimates it a little.

\begin{figure}
  \centering
  \includegraphics[width=.8\textwidth]{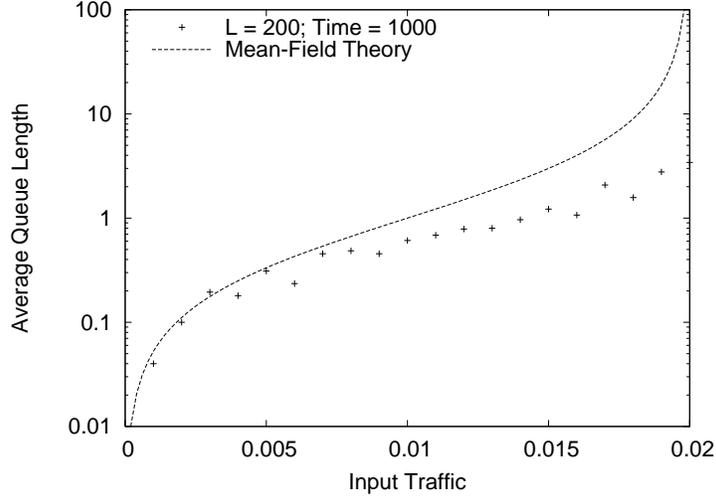}
  \caption{Comparison simulation to Mean-Field Theory
  on a one-dimensional sample with $L = 200, k = 1000$, for stable
  state queue length. The Mean-Field Fluid approximation
  overestimates the average queue length.}
  \label{profileQ1dL200Stable}
\end{figure}

The overestimation may originated from approximating the model to
an open Jackson network. Each site in the model is defined as an
M/D/1 queueing system, while it becomes to M/M/1 in the Jackson
network. It has been proved that, for Markovian routing schemes,
an M/D/1 queueing network has less average queueing delay (or,
equivalently, less queue length) in stable state than its M/M/1
counterpart \cite{harchol96net}.

\subsection{Fluid approximation of the Mean-Field Theory}

On the other hand, when the value of input traffic exceeds the
critical point, it is presented in the simulations that queue
length of a site approximately grows as a linear function to both
time and input traffic. To get this phenomenon explained, we
combine the assumptions of the Mean-Field Theory and
(\ref{eqn:MFTJackson:corollary:exitProb}) with the Fluid
approximation in queueing theory \cite{kleinrock75queueII}.

The Fluid approximation, based on the law of large numbers,
replaces discontinuous stochastic arrival and departure processes
with continuous deterministic versions. Let
$\overline{\alpha(\mathbf{x}, t)}$ and
$\overline{\delta(\mathbf{x}, t)}$ represents, respectively, the
two continuous processes for packet arrival and departure
happening at site $\mathbf{x} \in \mathcal{L}^d$ in our model.
Then for the queue length approximation $\overline{q(\mathbf{x},
t)}$, we have
\begin{equation}\label{eqn:fluidQ}
    \overline{q(\mathbf{x}, t)} = \overline{\alpha(\mathbf{x}, t)}
    - \overline{\delta(\mathbf{x}, t)}, \quad \forall \mathbf{x} \in
    \mathcal{L}^d
\end{equation}

Arrivals are resulted by both traffic input and forwarding events,
and the departure rate is constant $\mu = 1$. Therefore, we have
the Fluid version of (\ref{eqn:jacksonTheorem:sigma}).
\begin{eqnarray}\label{eqn:fluid}
    \overline{\alpha(\mathbf{x}, t)} &=& \overline{\alpha(\mathbf{x},
    0)} + \int_{0}^{t}\lambda dy + \sum_{\mathbf{y} \in
    A(\mathbf{x})}{r_{\mathbf{yx}}\overline{\delta(\mathbf{y},
    t)}}, \label{eqn:fluid:alpha}\\[.5em]
    \overline{\delta(\mathbf{x}, t)} &=& \overline{\delta(\mathbf{x},
    0)} + \int_{0}^{t} dy = \overline{\delta(\mathbf{x}, 0)} +
    t, \quad \forall \mathbf{x} \in \mathcal{L}^d
    \label{eqn:fluid:delta}
\end{eqnarray}

With (\ref{eqn:MFTJackson:corollary:exitProb}), i.e.
$r_{\mathbf{yx}} = (1 - 1/\bar\tau_0)/A$, another corollary is
obtained.
\begin{corollary}\label{theorem:MFTJackson:corollaryFluid}
    In the Mean-Field Theory of the model, approximated with Fluid model,
    queue length at any site is growing linearly if the input traffic
    exceeds the critical value, and the growth rate is
    \begin{equation}\label{eqn:fluidRate}
        \frac{d \overline{q}}{dt} = \lambda -
        \frac{1}{\bar\tau_0} = \lambda - \lambda_c, \quad \forall
        \mathbf{x} \in \mathcal{L}, t \rightarrow \infty
    \end{equation}
\end{corollary}

\begin{figure}
  \centering
  \subfigure[With fixed input traffic, the Mean-Field Fluid approach
  estimates that the slope of queue length to time is
  $\lambda - \lambda_c$, which is near to but a little less than the
  result of simulations.]{
    \label{fig:fluidMFTvsSimulation:1}
    \includegraphics[width=.8\textwidth]{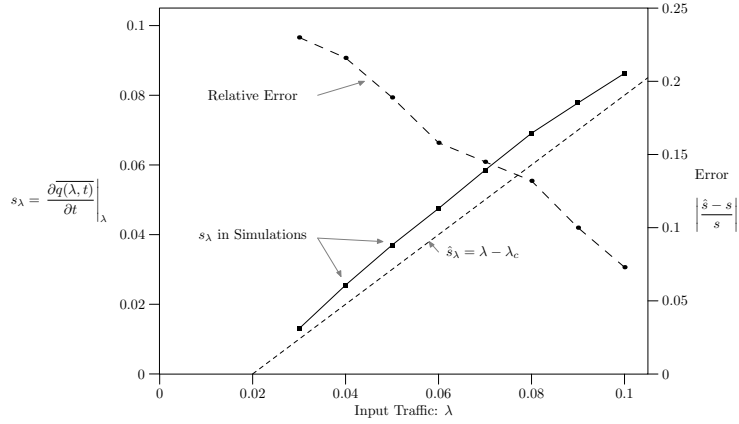}
  }
  \subfigure[With fixed time, the Mean-Field Fluid approach estimates
  that the slope of queue length to input traffic is $t$, which is
  near to but a little less than the simulation results too.]{
    \label{fig:fluidMFTvsSimulation:2}
    \includegraphics[width=.8\textwidth]{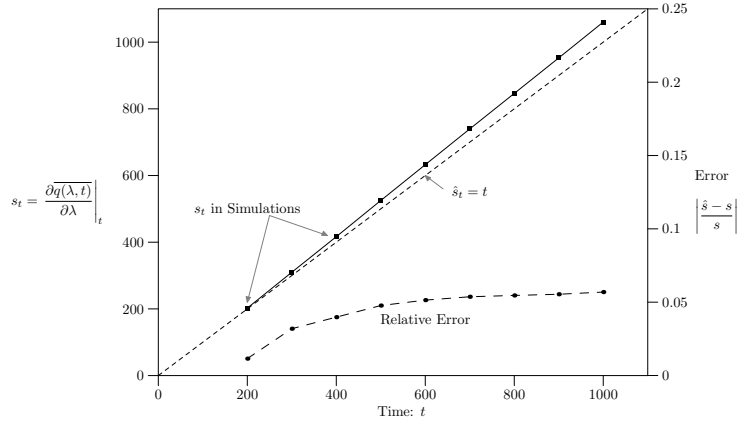}
  }
  \caption{Comparison the simulation results to Mean-Field Fluid
  approximation on one-dimensional samples with $L = 200$, for queue
  length growth far from stability. The Mean-Field Fluid approximation
  is near to but a little underestimates the queue growth rate.}
  \label{fig:fluidMFTvsSimulation}
\end{figure}

The result is compared to simulation in a way with either time
fixed and input traffic variant or vise-versa (Figure
\ref{fig:fluidMFTvsSimulation}). The comparison focuses on the
growth rate. It is reasonable that the Mean-Field Fluid
approximation a little underestimates the rate of queue growth
because Fluid approaches replace the stochastic processes with
deterministic (D/D/1) systems. The larger the input traffic is,
the more the rate of queue growth estimated with Mean-Field Fluid
approach is close to the simulation.

\section{Summary}

The work is a part of research efforts on distributed network
architectures. The property of being distributed makes a network
much more reliable but a little less efficient than a hierarchical
decentralized architecture, such as today's Internet. However, in
some circumstances, the network is so dynamically organized (such
as \emph{ad hoc} network) that reliability is far important rather
than the efficiency, even it is not possible to build any
centralized control or any hierarchy. In some other circumstances
such as metropolitan-area-networks, because the speed of physical
links has greatly increased than the age of early Internet,
distributed architecture may perform better in both reliability
and efficiency.

This paper develops a modelling method for studying distributed
network behaviors. The introduced Cellular Automata model can be
applied not only to analyzing but also to designing architectures,
where a topology like regular lattices is able to be implemented.
It promotes the previous works of simulation to an approximated
analytical result, where the law of critical traffic is proved
under a certain set of Mean-Field assumptions. Being advantageous
to finite simulations, the analytical result refines the
conditions of the law of critical traffic. Under these conditions,
i.e. isotropy, homogeneity and ergodicity, the critical traffic is
determined by the free delay of the network. Therefore, one may
improve the critical behavior of a network by improving the free
delay of its topology equivalently.

On the other hand, the Mean-Field Theory is not able to accurately
estimate some quantities such as average queue length in the
model. Generally speaking, the farther the input traffic is away
from the the critical point, the better the Mean-Field theory
estimates queue length behavior, no matter what kind of state the
model is in, stable or far from stability. In practice, quality
control protocol may predict the queueing tendency under either
definitely heavy or definitely light input and do actions in
response.

It is true that, once the assumptions are far from reality, the
Mean-Field Theory will be broken. For example, when some sites in
the model cease working, packets must bypass these sites and
therefore the isotropy of the model is destroyed. Since the
critical behavior depends upon the maximum site traffic, it could
be easily derived from (\ref{eqn:jacksonTheorem:sigma}) and has
been demonstrated in simulations that in such a case the critical
input traffic is significantly less than that of the Mean-Field
Theory, and improving free delay is still important because it
upper-bounds the maximum-possible critical traffic.

Quantitatively analyzing non-isotropic, non-homogeneous and
non-ergodic models, which are more close to real systems, are of
the successive work after this paper. Like studying any new
statistical-physical systems, one should get deep into the local
details of reaction, where the fluctuations are ignored by the
Mean-Field Theory.

\bibliographystyle{plain}
\bibliography{MFT4DistributedNet}

\end{document}